\documentstyle[aps,pra,epsfig,preprint]{revtex} 
\tightenlines

\def\be{\begin{equation}}
\def\ee{\end{equation}}
\def\bea{\begin{eqnarray}}
\def\eea{\end{eqnarray}}
\def\bma{\begin{mathletters}}
\def\ema{\end{mathletters}}
\def\C{\hbox{$\mit I$\kern-.6em$\mit C$}}

\begin{document}
\draft

\title{ The relative entropy of entanglement of Schmidt correlated states and distillation}

\author{Yi-Xin Chen and Dong Yang}

\address{Zhejiang Institute of Modern Physics and
Department of Physics, Zhejiang University, Hangzhou 310027, P.R. China} 

\date{\today}

\maketitle

\begin{abstract}
We show that a mixed state $\rho=\sum_{mn}a_{mn}|m\rangle \langle n|$ can be realized by an ensemble of pure states $\{p_{k}, |\phi_{k} \rangle \}$ where $|\phi_{k}\rangle=\sum_{m}\sqrt{a_{mm}}e^{i\theta_{m}^{k}}|m\rangle$. Employing this form, we discuss the relative entropy of entanglement of Schmidt correlated states. Also, we calculate the distillable entanglement of a class of mixed states.
\end{abstract}


\narrowtext

Quantum entanglement is at the heart of many aspects of quantum information theory and responsible for many quantum tasks such as teleportation \cite{Bennett1}, dense coding \cite{Bennett2}, quantum error correction \cite{Bennett3} etc. In this sense, it is nowadays viewed as a quantum resource. Intensive theoretical efforts were made to understand the mathematical structure of entanglement, qualitatively and quantitively. Among the entanglement measures, the relative entropy of entanglement \cite{VP1,VP2} plays an important role. For general mixed states, the relative entropy of entanglement is hard to calculate. However, it can be calculated explicitly for the Schmidt correlated states\cite{Rains,Wu} . In this note, employing the particular realization form of mixed states, we easily deduce the properties of Schmidt correlated states. Generally, orthogonal states may be distinguished perfectly only by means of global measurements since the global information of orthogonality may be encoded in entanglement that may not be extracted by local operations and classical communication (LOCC). Bennett et al. \cite{Bennett4} showed that a set of nine pairwise orthogonal product states in $3\otimes 3$ cannot be reliably distinguished by LOCC. Recently, Walgate et al. \cite{Walgate} demonstrated that any two orthogonal multipartite pure states could be distinguished with certainty by only LOCC operations. In general, more than two orthogonal entangled states cannot be discriminated if only LOCC operations are allowed. Ghosh et al. \cite{Ghosh} related the distinction process with the distillation one and showed that any three Bell states cannot be discriminated by LOCC operations. Also, they calculated the distillable entanglement of a certain class of mixed states whose distillable entanglement attains the relative entropy of entanglement. We will generalize their result to the Schmidt correlated states of high dimensions.

Before we discuss the Schmidt correlated states, we give a particular realization form of a mixed state, which is very helpful. 

{\it Lemma}: For a  general mixed state $\rho=\sum_{mn}a_{mn}|m\rangle \langle n|$, there exists an ensemble of pure states $\{p_{k}, |\phi_{k}\rangle\}$ realizing $\rho$, where $|\phi_{k}\rangle$ is of the form $|\phi_{k}\rangle=\sum_{m}\sqrt{a_{mm}}e^{i\theta_{m}^{k}}|m\rangle$.

Proof: If $\rho$ could be realized by the ensemble $\{p_{k}, |\phi_{k}\rangle\}$, then
\be
\sum_{k}p_{k}|\phi_{k}\rangle \langle \phi_{k}|=\sum_{mn}\sqrt{a_{mm}a_{nn}}\sum_{k}p_{k}e^{i(\theta_{m}^{k}-\theta_{n}^{k})}|m\rangle \langle n|=\sum_{mn}a_{mn}|m\rangle \langle n|.
\ee  
That is 
\be
\sqrt{a_{mm}a_{nn}}\sum_{k}p_{k}e^{i(\theta_{m}^{k}-\theta_{n}^{k})}=a_{mn},
\ee
should hold.
From the positivity of $\rho$, we know $\sqrt{a_{mm}a_{nn}}\ge |a_{mn}|$.
For $\rho$ is a $2\times 2$ matrix, it is sufficient to prove that there exists $\{p_{k}, \theta_{1}^{k}, \theta_{2}^{k}\}$ satisfying $\sqrt{a_{11}a_{22}}\sum_{k}p_{k}e^{i(\theta_{1}^{k}-\theta_{2}^{k})}=a_{12}$. Regard each term in the sum as a vector in the complex plane, it is easy to see that we can always choose $\{p_{k}, \theta_{12}^{k}\}$ satisfying $\sum_{k}p_{k}e^{i\theta_{12}^{k}}=\frac{a_{12}}{\sqrt{a_{11}a_{22}}}$. In fact, there exist infinite solutions. Note also that $p_{k}, k=1, \cdots, K$, where $K$ could be any large number. It is important for our induction.
Suppose that it is true for the case of $(L-1)\times (L-1)$ matrix. Then in the case of $L\times L$ matrix, $\rho_{L}=\sum_{mn}^{L}a_{mn}|m\rangle \langle n|$. From the positivity of $\rho_{L}$, $\tilde{\rho}_{L-1}=\sum_{mn}^{L-1}a_{mn}|m\rangle \langle n|$ is also positive and can be normalized as a density matrix $\rho_{L-1}=\frac{1}{N}\sum_{mn}^{L-1}a_{mn}|m\rangle \langle n|$, where $N=\sum_{m=1}^{L-1}a_{mm}$. According to the supposition, it can be realized by $\{p_{k}, |\phi_{k}\rangle\}$, where $|\phi_{k}\rangle=\frac{1}{\sqrt{N}}\sum_{m=1}^{L-1}\sqrt{a_{mm}}e^{i\theta_{m}^{k}}|m\rangle$. Now suppose $\rho_{L}$ could be expressed as $\rho_{L}=\sum_{k=1}^{K}p_{k}|\psi_{k}\rangle \langle \psi_{k}|$, where $|\psi_{k}\rangle$ are chosen as the following form
\be
|\psi_{k}\rangle=\sqrt{N}|\phi_{k}\rangle+\sqrt{a_{LL}}e^{i\theta_{L}^{k}}.
\ee
If there exist solutions for $\theta_{L}^{k}$, then the proof is completed. The $\theta_{L}^{k}$ must satisfy the following equations,
\be
\sum_{k=1}^{K}p_{k}e^{i(\theta_{m}^{k}-\theta_{L}^{k})}=\frac{a_{mL}}{\sqrt{a_{mm}a_{LL}}}, (m=1, \cdots, L). 
\ee  
It is clear that there always exist solutions of $\theta_{L}^{k}$ for $K > L$. And indeed, $K$ could be any large number. So the proof is completed. 

Employing the lemma, a particular realization form can be obtained for Schmidt correlated states. A bipartite mixed state $\rho$ is called as Schmidt correlated state if it can be expressed as $\rho=\sum_{mn}a_{mn}|mm\rangle \langle nn|$. Now we see immediately that Schmidt correlated state could be realized by an ensemble $\{p_{k}, |\phi_{k}\rangle\}$, where $|\phi_{k}\rangle=\sum_{m}\sqrt{a_{mm}}e^{i\theta_{m}^{k}}|mm\rangle$. Each $|\phi_{k}\rangle$ has the same Schmidt coefficients. Using this form, we discuss the properties of the Schmidt correlated state which were investigated in the papers \cite{Rains,Wu,Virmani}.

The relative entropy of entanglement\cite{VP1} for a bipartite quantum state $\rho$ is defined by 
\be
E_{r}(\rho)=min_{\sigma\in \cal{D}}S(\rho\|\sigma),
\ee
where $\cal{D}$ is the set of all separable states, and $S(\rho\|\sigma)=tr\rho(log\rho-log\sigma)$. Vedral and Plenio\cite{VP2} proved that $E_{r}(\rho)$ reduced to the von Neumann entropy of the reduced state of either side for pure bipartite state $|\phi_{k}\rangle=\sum_{m}\sqrt{a_{mm}}|mm\rangle$ and the optimal separable matrix $\sigma^{*}=\sum_{m}a_{mm}|mm\rangle\langle mm|$. Also, they showed that if $\sigma^{*}$ is optimal for $\rho$, then it is also optimal for $\rho^{'}=p_{1}\rho+p_{2}\sigma^{*}$. Now we extend their theorem further.

{\it Extended Vedral-Plenio theorem}: If $\rho_{i}$  have the same $\sigma^{*}\in \cal{D}$ which minimize $S(\rho_{i}\|\sigma^{*})$, then $\sigma^{*}$ is also the optimal operator for $\rho=\sum_{i}p_{i}\rho_{i}$.

Proof:
From the definition of $S(\rho\|\sigma)$, it is sufficient to minimize $-tr\rho log\sigma=-\sum_{i}p_{i}tr\rho_{i}log\sigma$.  
As $\sigma^{*}$ is optimal for $\rho_{i}$, we know
\be
-tr\rho_{i}log\sigma^{*}=\inf_{\sigma\in \cal{D}}(-tr\rho_{i}log\sigma)\le -tr\rho_{i}log\sigma .
\ee
So $\sigma^{*}$ is optimal for $\rho$. It is clear to see the physical meaning of $E_{r}(\rho)$ after explicitly writing the expression of $E_{r}(\rho)$  
\be
E_{r}(\sum_{i}p_{i}\rho_{i})=\sum_{i}p_{i}S(\rho_{i}\|\sigma^{*})-\sum_{i}p_{i}S(\rho_{i}\|\rho).
\ee
The first term is the average quantum entanglement, while the second term is the lost classical information caused by mixing process. From the particular realization form of Schmidt correlated state and the extended Vedral-Plenio theorem, we immediately calculate the relative entropy of entanglement of Schmidt correlated states.

{\it Corollary 1}: For Schmidt correlated state $\rho=\sum_{mn}a_{mn}|mm\rangle \langle nn|$, the relative entropy of entanglement $E_{r}(\rho)=S(\rho\|\sigma^{*})$, where $\sigma^{*}=\sum_{m}a_{mm}|mm\rangle \langle mm|$.

Note that the same result has been obtained by Rains\cite{Rains} and Wu et al\cite{Wu}. Furthermore, we investigate the additivity of the relative entropy of entanglement of Schmidt correlated states. 

{\it Corollary 2}: The relative entropy of entanglement for Schmidt correlated states are additive, i. e. for any two Schmidt correlated states $\rho_{1}, \rho_{2}$, $E_{r}(\rho_{1}\otimes \rho_{2})=E_{r}(\rho_{1})+E_{r}(\rho_{2})$.
Proof:
The Schmidt correlated states $\rho_{1}$ and $\rho_{2}$ can be realized by $\{p_{i}, |\phi_{i}\rangle\}$ and $\{q_{j}, |\psi_{j}\rangle\}$ respectively. 
\be
|\phi_{i}\rangle=\sum_{m}\sqrt{a_{mm}}e^{i\theta_{m}^{i}}|mm\rangle, \nonumber\\
|\psi_{j}\rangle=\sum_{n}\sqrt{b_{nn}}e^{i\theta_{n}^{j}}|nn\rangle,
\ee  
 where $a_{mm}$ and $b_{nn}$ are the diagonal elements of $\rho_{1}$ and $\rho_{2}$, $|m\rangle, |n\rangle$ are two basis and not necessarily the same. All the $|\phi_{i}\rangle$ have the same optimal separable state $\sigma_{1}^{*}=\sum_{m}a_{mm}|mm\rangle\langle mm|$ as $\rho_{1}$ and$|\psi_{i}\rangle$ have $\sigma_{2}^{*}=\sum_{n}a_{nn}|nn\rangle\langle nn|$ as $\rho_{2}$.
 \be
\rho_{1}\otimes \rho_{2}=\sum_{ij}p_{i}q_{j}|\phi_{i}\rangle\langle\phi_{i}|\otimes|\psi_{j}\rangle\langle\psi_{j}|.
\ee
$\rho_{1}\otimes \rho_{2}$ is also a Schmidt correlated state. And $\sigma_{1}^{*}\otimes\sigma_{2}^{*}$ is the optimal separable state for the product pure states $|\phi_{i}\rangle|\psi_{j}\rangle$, so it is the optimal one for $\rho_{1}\otimes \rho_{2}$.
\bea
E_{r}(\rho_{1}\otimes \rho_{2})&=&S(\rho_{1}\otimes \rho_{2}\|\sigma_{1}^{*}\otimes\sigma_{2}^{*}),\nonumber\\
&=&tr\rho_{1}\otimes \rho_{2}log\rho_{1}\otimes \rho_{2}-\sum_{ij}p_{i}q_{j}|\phi_{i}\rangle\langle\phi_{i}|\otimes|\psi_{j}\rangle\langle\psi_{j}|log\sigma_{1}^{*}\otimes\sigma_{2}^{*}, \nonumber\\
&=&tr\rho_{1}(log\rho_{1}-log\sigma_{1}^{*})+tr\rho_{2}(log\rho_{2}-log\sigma_{2}^{*}), \nonumber\\
&=&E_{r}(\rho_{1})+E_{r}(\rho_{2}).
\eea
So the relative entropy of entanglement is additive for Schmidt correlated states. The conclusion has been proved by Rains\cite{Rains}. 

The subspace spanned by all the states in the range of any Schmidt correlated state $\rho$ is called as Schmidt correlated subspace of the state $\rho$. It is clear that all the pure  states in the Schmidt correlated subspace have the same Schmidt basis in the Schmidt decomposition form. A set of pure states is called Schmidt correlated if they lie in the same Schmidt correlated subspace. Virmani et al. \cite{Virmani} discussed the notion of Schmidt correlated pure states, which was introduced by Rains as maximal correlated states. A set of Schmidt correlated pure states can always be discriminated locally as well as globally, regardless of which figures of merit are chosen. They investigated the conditions under which two pure states can be written in Schmidt correlated form and showed that any two maximally entangled states can always be expressed in Schmidt correlated form, thus showing that two maximally entangled states can always be discriminated locally according to any figures of merits. In the following, we will discuss a class of mixed states whose distillable entanglement can be calculated explicitly. This generalizes the result of Ghosh et al. In \cite{Ghosh}, the distillable entanglement of the mixed state of the form 
\be
\rho=\frac{1}{2}(|\Phi_{i}\rangle^{\otimes 2}\langle\Phi_{i}|+|\Phi_{j}\rangle^{\otimes 2}\langle\Phi_{j}|),
\ee
is shown to be one ebit, where $|\Phi_{i}\rangle, |\Phi_{j}\rangle$ are two different Bell states. We know any two Bell states are Schmidt correlated. If a set of Schmidt correlated pure states are orthogonal, they can be distinguished perfectly by local operations and classical communication even if they are entangled. We give the example in $3\times 3$ and it is easy to generalize to high dimensions. Three orthogonal states in the Schmidt correlated subspace $V$ spanned by $\{|00\rangle, |11\rangle, |22\rangle\}$ are generally of the form:
\be
|e_{i}\rangle=\sum_{j}u_{ij}|jj\rangle, i,j=0,1,2
\ee
where $u_{ij}$ are the elements of a unitary matrix. Under another basis $\{|i^{'}\rangle\}$ 
\bea
|0\rangle&=&|0^{'}\rangle+|1^{'}\rangle+|2^{'}\rangle, \nonumber\\ |1\rangle&=&|0^{'}\rangle+e^{i\frac{2}{3}\pi}|1^{'}\rangle+e^{i\frac{4}{3}\pi}|2^{'}\rangle,\nonumber \\ 
|2\rangle&=&|0^{'}\rangle+e^{i\frac{4}{3}\pi}|1^{'}\rangle+e^{i\frac{2}{3}\pi}|2^{'}\rangle,
\eea
$|e_{i}\rangle$ can be expressed as 
\bea
|e_{i}\rangle=|0^{'}\rangle(u_{i0}|0\rangle+u_{i1}|1\rangle+u_{i2}|2\rangle)+|1^{'}\rangle(u_{i0}|0\rangle+u_{i1}e^{i\frac{2}{3}\pi}|1\rangle+u_{i2}e^{i\frac{4}{3}\pi}|2\rangle) \nonumber\\
+|2^{'}\rangle(u_{i0}|0\rangle+u_{i1}e^{i\frac{4}{3}\pi}|1\rangle+u_{i2}e^{i\frac{2}{3}\pi}|2\rangle), i=0,1,2.
\eea
The first party performs measurement in the basis $\{|i^{'}\rangle\}$, any outcome would project the second side onto an orthogonal basis. So the second party does measurement in the corresponding basis and tells which the state is. Now we calculate the distillable entanglement of a class of mixed states that are classical correlated between two sets of Schmidt correlated pure states, 
\be
\rho_{A{1}A_{2}B_{1}B{2}}=\frac{1}{N}\sum_{i=1}^{N}|e_{i}\rangle_{A_{1}B_{1}}\langle e_{i}|\otimes|\phi_{i}\rangle_{A_{2}B_{2}}\langle \phi_{i}|,
\ee 
where
$|e_{i}\rangle=\sum_{j}u_{ij}|jj\rangle$ 
are an orthogonal basis in a Schmidt correlated subspace and 
$|\phi_{i}\rangle=\sum_{k}\sqrt{\lambda_{k}}e^{i\theta_{k}^{i}}|kk\rangle$.
All the $|\phi_{i}\rangle$ have the same optimal separable state $\sigma^{*}=\sum_{k}\lambda_{k}|kk\rangle\langle kk|$.
Employing the distinction process for the distillation protocol, the distillable entanglement of $\rho$ in the bipartite cut of $A_{1}A_{2}:B_{1}B_{2}$ is at least $S(|\phi\rangle)$, where $S(|\phi\rangle)=-\sum p_{i}\log p_{i}$ is the entanglement of the pure state $|\phi\rangle$. So we have an inequality
\be
E_{d}(\rho)\ge S(|\phi\rangle)=-\sum_{i}\lambda_{i}log\lambda_{i}.
\ee
The relative entropy of entanglement of $\rho$ can be explicitly calculated since $\rho$ is a Schmidt correlated state.
\be
E_{r}(\rho)= S(|\phi\rangle)=-\sum_{i}\lambda_{i}log\lambda_{i}.
\ee  
We know that the relative entropy of entanglement is an upper bound on the distillable entanglement, that is 
\be
E_{r}(\rho)\ge E_{d}(\rho).
\ee
So $E_{d}(\rho)=E_{r}(\rho)$ is obtained. The same reasoning has also been employed in \cite{Ghosh}. Notice that in order to discriminate the orthogonal pure states in the Schmidt correlated subspace, the entanglement has to be destroyed completely. This gives us a clue for calculating the distillable entanglement of the general Schmidt correlated state. We will investigate this problem further.

In summary, we find a particular realization form for a mixed state and employ it to discuss the relative entropy of entanglement of Schmidt correlated state. Also we provide a class of mixed states whose distillable entanglement can be calculated.

D. Yang thanks S. J. Gu and H. W. Wang for helpful discussion. The work is supported by the NNSF of China, the Special NSF of Zhejiang Province (Grant No.RC98022) and Guang-Biao Cao Foundation in Zhejiang University.



\begin{references}

\bibitem{Bennett1}
C. H. Bennet, G. Brassard, C. Crepeau, R. Jozsa, A. Peres and W. K. Wootters, Phys. Rev. Lett {\bf 70}, 1895 (1993).

\bibitem{Bennett2}
C. H. Bennett and S. J. Wiesner, Phys. Rev. Lett {\bf 69}, 2881(1992).

\bibitem{Bennett3}
C. H. Bennett, D. P. DiVincenzo, J. A. Smolin and W. K. Wootters, Phys. Rev. A {\bf 54}, 3824 (1996).

\bibitem{VP1}
V. Vedral, M. B. Plenio, M. A. Rippin, and P. L. Knight, Phys. Rev. Lett {\bf 78}, 2275 (1997).

\bibitem{VP2}
V. Vedral and M. B. Plenio, Phys. Rev. A {\bf 57}, 1619 (1998).

\bibitem{Rains} 
E. M. Rains, Phys. Rev. A {\bf 60}, 179(1999).

\bibitem{Wu}
S. Wu and Y. Zhang, Calculating the relative entropy of entanglement, quant-ph/0004020.

\bibitem{Bennett4}
C. H. Bennett, D. P. DiVincenzo, C. A. Fuchs, T. Mor, E. Rain, P. W. Shor, and
 J. A. Smolin, Phys. Rev. A {\bf 59}, 1070 (1999).

\bibitem{Walgate}
J. Walgate, A. Short, L. Hardy and V. Vedral, Phys. Rev. Lett. {\bf 85}, 4972 (2000).

\bibitem{Ghosh}
S. Ghosh, G. Kar, A. Roy, A. Sen(De) and U. Sen, Phys. Rev. Lett. {\bf 87}, 277902 (2001), quant-ph/0106148.

\bibitem{Virmani}
S. Virmani, M. F. Sacchi, M. B. Plenio and D. Markham, Phys. Lett. A {\bf 288}, 62 (2001).

\end{references}
\end{document}